\newcommand{\yr}{\,{\rm yr}}
\newcommand{\masyr}{\,{\rm mas}\yr^{-1}}
\newcommand{\simle}{\mbox{$\stackrel{<}{_{\sim}}$}}
\newcommand{\simge}{\mbox{$\stackrel{>}{_{\sim}}$}}
\def\arcdeg{\hbox{$^\circ$} }
\def\arcdegg{\hbox{$^\circ$}}
\newcommand{\micron}{\,\hbox{$\mu$m} }

\def\etal{ {\em et~al.\/}\thinspace}
\def\arcsec{\hbox{$^{\prime\prime}$} }

\def\msun{\hbox{\,M$_\odot$}}
\def\rsun{\hbox{\,R$_\odot$}}

\def\kms{ km~s$^{-1}$}

\documentstyle[epsfig,emulateapj]{article}
\lefthead{Monnier et al.}
\righthead{Pinwheel nebula around WR~98a}

\begin{document}
\title{Pinwheel Nebula around WR~98a}
\author{J. D. Monnier\altaffilmark{1}, P. G. Tuthill\altaffilmark{2}, 
and W. C. Danchi}

\altaffiltext{1}{Current Address: Smithsonian Astrophysical Observatory MS\#42,
60 Garden Street, Cambridge, MA, 02138}
\altaffiltext{2}{Current Address: Chatterton Astronomy Dept, 
School of Physics, University of Sydney, NSW 2006, Australia}

\affil{Space Sciences Laboratory, University of California, Berkeley,
Berkeley,  CA  94720-7450}

%
%
%
\begin{abstract}
We present the first near-infrared images of dusty Wolf-Rayet star
WR~98a.  Aperture masking interferometry has been utilized to recover
images at 
the diffraction-limit of the Keck-I telescope, $\simle$50\,mas
at 2.2\micron.  Multi-epoch observations spanning about
one year have resolved the dust shell into a ``pinwheel'' nebula, the
second example of a new class of dust shell first discovered 
around WR 104
(Tuthill, Monnier, \& Danchi 1999a). Interpreting the
collimated dust outflow in terms of an interacting winds model, 
the binary orbital parameters and apparent
wind speed are derived: a period of $565\pm 50$ days, a viewing angle
of $35\arcdegg\pm 6$\arcdeg\, from the pole, and a 
wind speed of $99\pm 23\masyr$.  This
period is consistent with a possible $\sim$588\,day periodicity in the
infrared light curve (\cite{williams95}), linking the photometric
variation to the binary orbit.  Important implications for binary
stellar evolution are discussed by identifying WR~104 and WR~98a as
members of a class of massive, short-period binaries whose orbits
were circularized during a previous red supergiant phase.  The current
component separation in each system is similar to the diameter of a
red supergiant, indicating that the supergiant phase was likely
terminated by Roche-lobe overflow, leading to the present Wolf-Rayet
stage.

\end{abstract}


\section{Introduction}
A small fraction of Wolf-Rayet (WR) stars are known to be strong
infrared (IR) sources, surrounded by shells of warm dust maintained by
massive stellar winds.  These systems have been classified as either
``variable'' or ``persistent'' dust-producers, based on the
variability of IR flux (\cite{wh92}).  Radio
observations of some ``variable'' WR stars have detected non-thermal
emission, interpreted as the interface between colliding winds in {\it
long-period} WR binary systems (e.g., \cite{wb95}; \cite{veen98};
\cite{williams98}).
The episodic formation of dust in these systems appears to coincide
with periastron passage,
the colliding winds at close binary separation (a
few AU) apparently catalyzing dust formation.  
The radio and IR
properties of the prototypical system WR\,140 and other episodic
dust producers have been successfully explained in the context of such
wind-wind models (e.g., \cite{moffat87}; \cite{williams90}).

Most of the brightest infrared WRs, however, are classified as
``persistent'' dust producers, and these objects have been more
difficult to explain.
The lack of IR variability suggests that dust is being
produced constantly, despite the unfavorable dust formation conditions
around single WR stars (\cite{ct95}).  This has led
to the suggestion that {\em short-period} binaries lie buried in these
optically-obscured systems with wind-wind collisions catalyzing the
dust formation (see \cite{usov91};\cite{wh92}).  
However, other workers have argued that enough dust can
form in the spherically-symmetric wind of a singular WR via novel dust
formation processes (e.g., \cite{zubko98}).  Until recently, few
direct observations were available to settle this controversy.

A diffraction-limited, multi-epoch study of these IR-bright sources at
near-IR wavelengths is underway using the Keck-I telescope.  Initial
results for the IR-bright WR~104 (WC9 [\cite{tcm86}]) have been
published in Tuthill\etal (1999a), marking the discovery of the first
``pinwheel'' nebula.  We now present the second such nebula WR~98a,
firmly establishing this new class of dust shell.  Although one of the
IR-brightest WR stars known, WR~98a (WC8-9 [\cite{williams95}]) was only
recently identified through the use of IRAS data (\cite{cohen91}), and no 
previous imaging of its dust shell has been reported.
While no variability has been detected for WR~104, $\sim$0.5~mag
(K-band) variability has been observed for WR~98a (\cite{williams95}).
For this reason, it is often included as a ``variable'' dust producer,
although this level of variability is significantly less than that
observed for WR~140 and other long-period binary systems.

\section{Observations}
Aperture masking interferometry was performed by placing aluminum masks 
in front of the Keck-I infrared 
secondary mirror.  This technique converts the primary
mirror into a VLA-style interferometric array, allowing the Fourier
amplitudes and closure phases for a range of baselines to be recovered with
minimal ``redundancy'' noise (e.g., \cite{baldwin86}).
The Maximum Entropy Method (MEM)
(\cite{gs84}; \cite{sivia87}) has been used to reconstruct 
diffraction-limited 
images from the interferometric data.  In order to check the
reliability of the reconstructions, the MEM results have been compared with
those from the CLEAN reconstruction algorithm (\cite{hogbom74}).
Further engineering and performance details may be found in
Monnier\etal (1999), Tuthill\etal(1999b), and Monnier (1999).

WR~98a was observed in June and September of 1998 and April 1999 at Keck-I
using the Near Infrared-Camera  (\cite{ms94}; \cite{matthews96}) and 
an annulus aperture mask.  Spectral filter characteristics and the number of
speckle frames (integration time 0.137\,s) obtained for each observation  
can be found in Table~1.
The unresolved star HD 163428 (spectral type K5II) was used
to calibrate the atmosphere plus telescope transfer function.

Multi-epoch images of WR~98a appear in Figure\,\ref{fig:wr98a}; the high
spatial resolution of the Keck ($\simle$50\,mas) was adequate to
resolve much of the dust shell around this source.  As for WR~104, the dust
emission was observed to be distributed in a rotating ``pinwheel'' nebula,
naturally explained by wind interactions between the WR and an OB-type
companion in a {\it short-period} binary system.  By inspection, the observed
rotational period is 1-2\,yr, corresponding to an orbital major axis of a
few AU for a massive binary system.  This separation amounts to
only a
few milli-arcseconds on the sky, unresolvable by this experiment.

\medskip
\vbox{
\epsfig{file=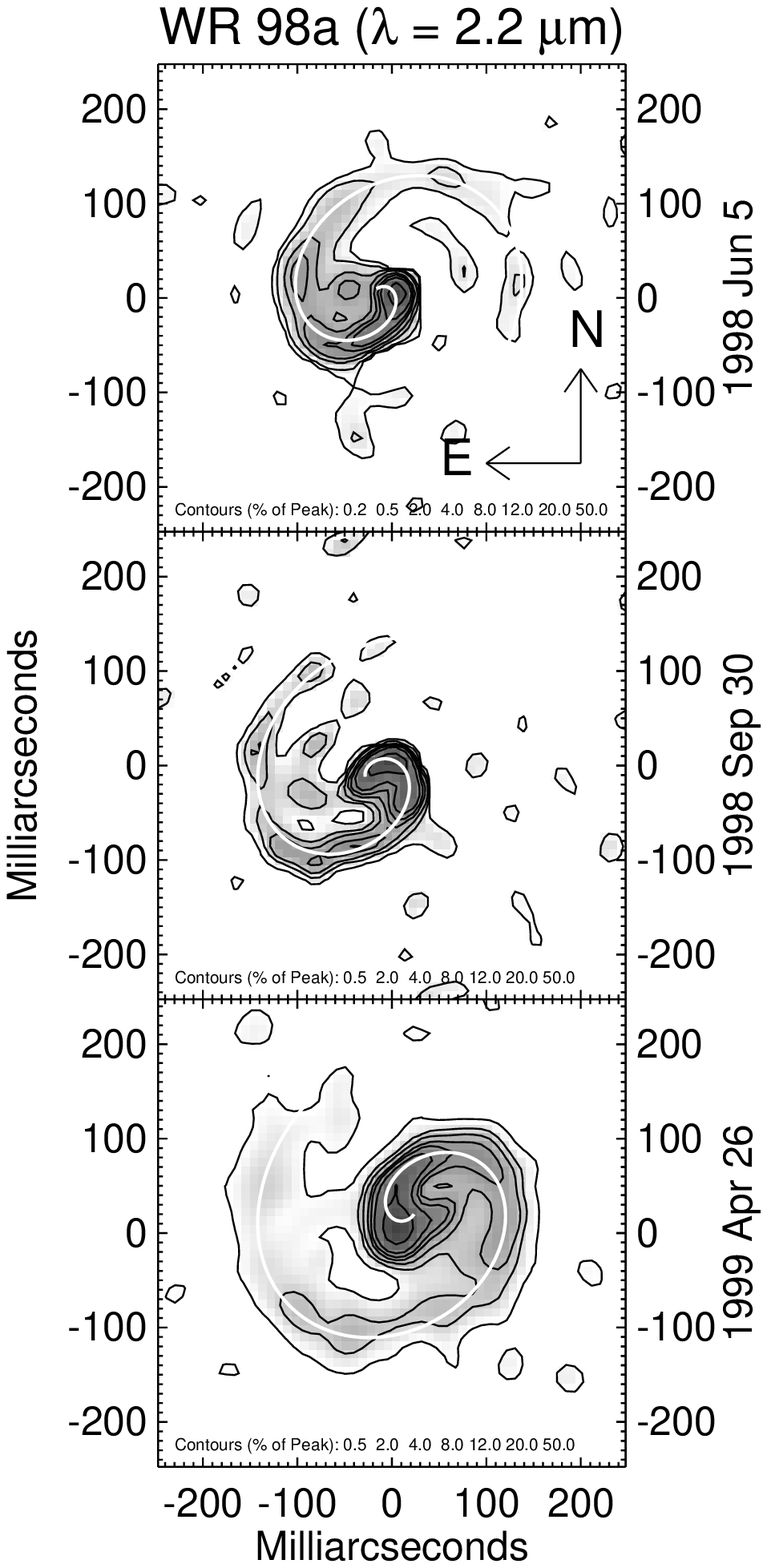,width=3.1in}
\smallskip
{

\noindent
\footnotesize Fig 1.--- Three~epochs of the 2.2\,$\mu$m emission from WR~98a
clearly show a spiral morphology.
The solid line represents the best fit plume morphology based on a simple
model (see \S\ref{section:spiral_model} \& \ref{section:results}).
\label{fig:wr98a}}}

\vbox{
\begin{center}
Table 1. --- Observing log for WR 98a at Keck-I
\epsfig{file=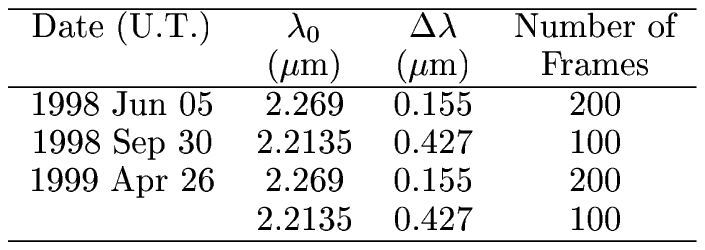,width=3.2in}
\end{center}}
\section{The Emission Model}
\label{section:spiral_model}
In the case of a WR+OB binary, both stars drive their own wind,
but the
momentum in the WR wind is expected to be significantly higher than
for the companion wind (e.g., by a factor of $\sim60$ for WR\,146
[\cite{dougherty96}]).  Hence, a shock will form between
the stars much closer to the OB companion, and the colliding gas will
be compressed and ultimately flow out of the interface region in
a wake behind the OB-star.  Although the detailed mechanisms are not known,
dust formation is catalyzed in such systems (\cite{moffat87}; 
\cite{williams90}; \cite{wh92}).
Generally, it is thought that dust can form in the compressed wake 
when the gas cools sufficiently.  High gas density may not be the only
important condition, as recent work indicates that the presence of hydrogen
from the OB companion wind may play an important dynamical 
role in efficient dust formation
(\cite{leteuff99}). 

Regardless of the dust formation mechanism, 
the highly collimated appearance
of the spirals in the WR~104 and WR~98a suggest that wind momenta
ratios $\big(\dot{M}v_\infty\big)_{\rm WC} / \big(\dot{M}v_\infty\big)_{\rm O}$
in these stars are indeed quite high, perhaps $\simge$100 (Canto, Raga \&
Wilkin 1996).  A more
quantitative discussion concerning the flow of cold gas from the 
wind-wind collision region can be found in Usov (1991).

As was done for WR~104 (\cite{tuthill99a}), 
the multi-epoch morphology of the WR~98a spiral can
be used to extract parameters of the obscured binary system,
including the period, inclination, and wind speed.  Our
model assumes that the dust forms in the wind moving at
constant (terminal) velocity.  In this case, the dust in the spiral is not
flowing {\em along} the spiral shape, but rather the dust is flowing
purely radially away from the binary; the spiral only appears to rotate in time
due
to the rotating dust formation site associated with the OB-companion.
A schematic diagram of the suggested binary geometry can be found in
Tuthill\etal(1999a).

Mathematically, the outflow appears as an Archimedean spiral; in polar
coordinates, $r=\alpha\theta$.  The curvature of the spiral is
controlled by the product of the wind speed and the period, which
corresponds to the distance the
dust travels during one period of the orbit.  
The period is determined from the apparent rotation manifest over
multi-epoch data, and together with the curvature of the spiral, leads
to a determination of the dust outflow velocity.
Lastly, the Archimedean spiral shape must be
projected onto the sky at some viewing angle.  Circularity of
the binary orbit was assumed (i.e. constant angular velocity)
and led to an excellent
simultaneous fit to the spiral morphology at all epochs, implying that the
orbit is not highly eccentric.
The current dataset is limited to
only three epochs and does not justify the additional model complexity
necessary
to precisely estimate the orbital eccentricity.  

\section{Results}
\label{section:results}
For WR~98a, fits were done to the combined three-epoch dataset by sampling the
K-band emission from the central bright core out to about 2\% of the peak,
resulting in the model curves of 
Figure\,\ref{fig:wr98a}.
The period was found to be 565$\pm$50 days, the viewing angle
35\arcdegg$\pm$6\arcdegg, and the angular velocity of the dust
99$\pm$23 mas\,yr$^-1$
(assuming a 565\,day period); the uncertainties in these parameters were
estimated as discussed below.  Using a crude estimate of the outflow
velocity, $\sim900$\kms based on near-IR linewidths
(\cite{williams95}), the angular velocity can be converted into
a distance estimate of $\sim$1.9\,kpc
(compare to 3$\pm$1\,kpc by \cite{cohen91}).  
The period, outflow velocity, and estimated distance are similar to 
those obtained for the other known pinwheel nebula WR 104:
220$\pm$30~days, 111$\pm$17~ mas/yr, and 2.3$\pm$0.7\,kpc respectively
(\cite{tuthill99a}).
However, the viewing angle of the WR~104 binary is significantly
closer to the pole, 20\arcdeg$\pm$5\arcdegg.

The uncertainties in the model parameters were estimated by
determining the span which yielded fits within 
35\% of the minimum $\chi^2$, and the suitability of this
criterion was checked visually by inspecting fits with a variety
of $\chi^2$.
Because the morphology of the spiral
curves is not greatly affected by the limited random noise in the maps,
potential {\em systematic} errors (e.g., optical depth effects, 
limited spatial resolution, low-level MEM artifacts)  
are the primary causes of mis-estimation in
model parameters.  

Faint structures in the WR~98a maps ($\sim$2\% of peak) are present
which do not appear to be part of the overall spiral morphology.  In
the June 1998 map, a spot or low-level extension appears east
of the peak.  Likewise in September 1998, low-level emission appears
between the bright core and the outer spiral.  
Such additional emission was not present
in WR~104 maps at any epoch, and requires an explanation beyond
the simple outflow model developed in \S\ref{section:spiral_model}.
While it is possible mapping artifacts are responsible for the
apparent emission, the structures are present in all data for a given
observing run and appear in CLEAN image reconstructions as well. 
We are left to conclude
that the structures are probably real, but we cannot presently explain them.

\section{Discussion}
\label{section:discussion}
The 1.55\,yr (565 day) orbital period is in good agreement with the
$\sim$1.4\,yr photometric period found by Williams\etal (1995).  In
Figure\,\ref{fig:wr98a_lightcurve}, K-band photometry from this paper
has been plotted along with a pessimistic estimate of the uncertainty.
In order to more precisely compare the possible periodicity in the IR
flux with the estimated period of the binary, a sinusoid was fit to
this data (see Figure\,\ref{fig:wr98a_lightcurve}) with best-fit
parameters as follows: the mean magnitude is 4.52, the peak-to-trough
amplitude is 0.65~mag, the date of maximum was 1992.59 and the period
is 1.61~yrs (588~days).

Since this photometry only sparsely samples a 2.4 yr span, the 1.61 yr
periodicity is difficult to firmly establish.  However, this
photometric period of 588 days is within 
errors of the binary orbital
period of 565$\pm$50 days derived in the last section.
This unlikely
coincidence suggests that 
\vbox{
\epsfig{file=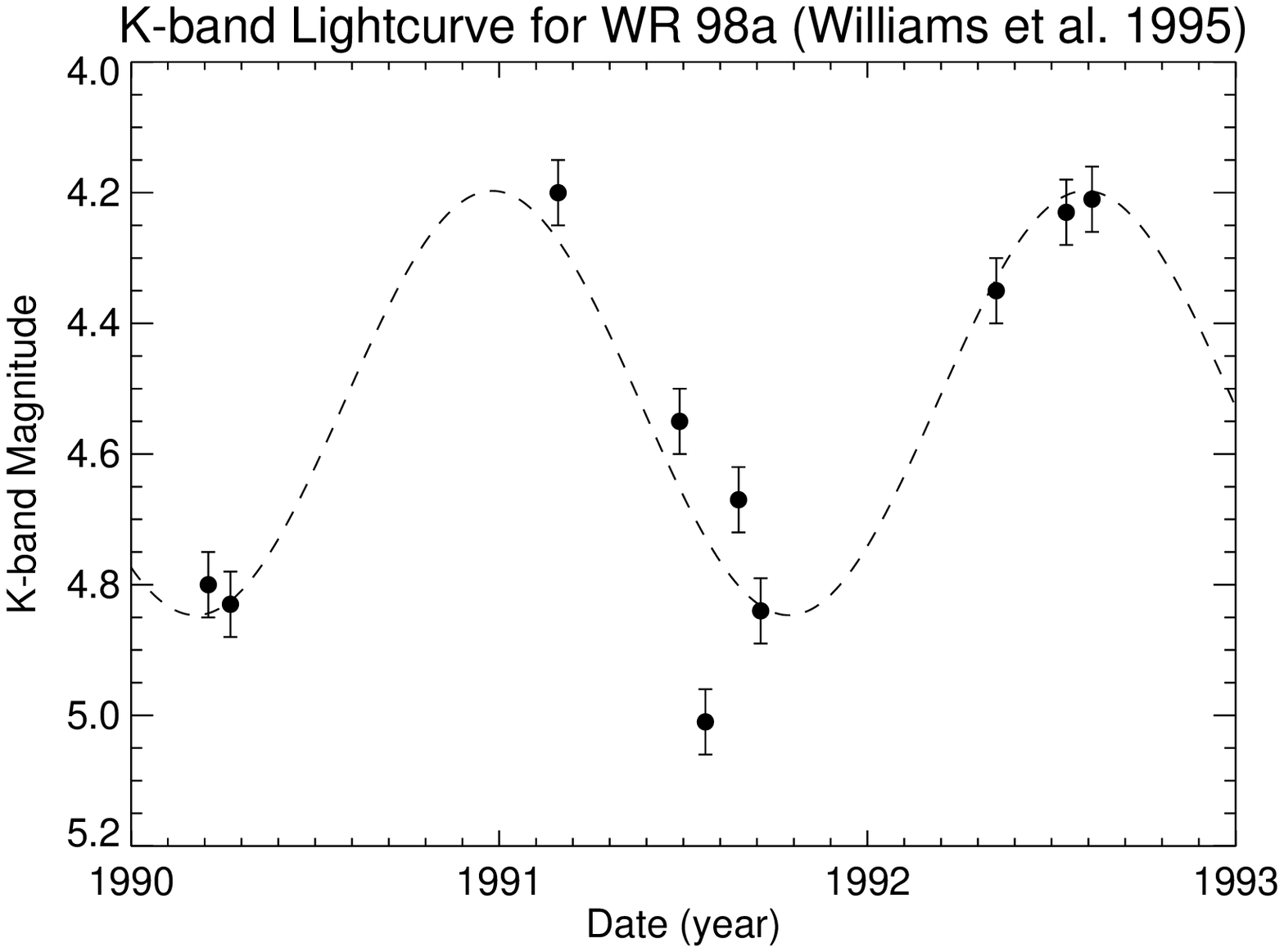,width=3.2in}

\noindent
\footnotesize Fig 2.---
K-band photometry of WR~98a, published in
Williams\etal(1995), shows evidence of variation.
The dashed line represents a best-fit sinusoid and the parameters are
reported in the text (\S\ref{section:discussion}).
\label{fig:wr98a_lightcurve}
\medskip}
the variation in the IR flux is directly
related to the binary itself, suggesting two possible scenarios.  The
0.65~mag variation may be due to the relatively high (35\arcdegg)
viewing angle from the pole, varying optical depth and perspective
could modulate the IR flux during the orbit.
Alternatively, ellipticity in the binary orbit could cause the dust
production rate to vary in time, causing the observed variation in IR
(i.e. thermal) radiation.  If the latter proves true, the eccentricity
is likely to be small, based on the relatively small IR variation and
the consistency of the three-epoch 
(circular) fit presented in
\S\ref{section:results}.  Recent unpublished photometry indicates that
WR~98a was in the middle of its rising light curve in June 1998, near
maximum light in September 1998, and fading in April 1999
(\cite{williamsPC}).  Better sampling of the spiral emission
morphology with coeval photometry through an orbital period 
is needed to determine the cause of IR variability and to 
precisely estimate the orbital eccentricity.
In light of this, precision photometry of WR~104 should also be 
obtained in order to search for a similar flux modulation at the orbital
period of its binary system (most recent period determination 
is 243.5$\pm 3.0$~days [Monnier 1999]).

\section{Implications for Binary Evolution}
Wolf-Rayet binary systems have been seen with periods $\simge$10\,yrs
and eccentric orbits which facilitate periodic dust production near
periastron passage (WR 140 [\cite{moffat87};
\cite{williams90}], WR 19 [\cite{veen98}], WR 137
[\cite{marchenko99}]).  However, the periods of the {\em
short-period} binaries reported around WR~104 and WR~98a
are $\sim$1\,yr and the orbits
appear nearly circular.  Circularity of WR~104 is supported by 
both the lack of IR variability, indicating a constant dust production
rate, and recent fits to four-epoch data under the assumption
of circularity (Monnier [1999]).  The level of 
ellipticity in the WR~98a orbit is not
known, but fits to three epochs of data (presented here) 
and the small IR variability
(as compared to long-period systems such as WR~140) do 
suggest the orbital eccentricity is small.  
Interestingly, the longest period WR~+~O system with a circular orbit
previously known is of order 30 days (Tassoul 1990).
Because of the unlikelihood of {\em a priori} circular orbits for
non-interacting binaries, the possibility that the orbits of WR~104
and WR~98a were {\em circularized} must be considered.  For massive
(M$\simge$10\,\msun) stars, a $\sim$1\,yr binary period corresponds to
a few AU in physical separation, while WR and O stars have radii of at
most a few \rsun.  The efficiency of tidal circularization is quite
sensitive to these parameters, and theory predicts no significant
orbital circularization under these circumstances (\cite{tassoul90};
\cite{verbunt95}).  The observed circularity can be explained,
however, if one of the stars was much larger in the past.

A WR star is characterized 
by the lack of a significant hydrogen envelope, lost
during previous phases of evolution, especially the red supergiant phase
(e.g., \cite{vanbeveren98}).  The diameter of a late-type
red supergiant can easily exceed an AU (e.g., \cite{vanbelle99}),
and hence tidal forces and subsequent circularization effects must 
have been important at this evolutionary stage.  The fact that the 
current component separation is about
that of a red supergiant diameter (or smaller!) suggests that binary
interaction, such as Roche lobe overflow, likely terminated the red
supergiant phase, ejecting a large fraction of the hydrogen envelope, and
leading to the onset of the
WR-stage in the primary (Vanbeveren\etal [1998]).  A
large ejection of material could explain the high optical
extinction to WR 104 (\cite{cohen75}),
despite a near pole-on viewing angle of the binary
system.  Similarly, heavy obscuration of a star only
$\sim$3\arcsec away from WR~98a (\cite{cohen91};
\cite{williams95}) suggests the local presence of a large amount
of unseen material.

If indeed persistent dust producing WR systems do imply the presence
of a short-period binary, then we are left to explain the observed
correlation of WR spectral type with the presence of dust; dust is
preferentially observed around WC8-9 stars (\cite{williams87}).
These late-type WC stars are distinguished from the early-type
through their wind ionization structure (\cite{tcm86}), which is not,
by itself, a clear signature of previous binary interactions.  One
explanation is that the enhanced mass-loss associated with the
formation of circular, $\sim$1\,yr binaries may lead to WR stars with
late-WC properties, i.e. lower terminal velocities and less-excited
wind ionization structure.  The high mass-loss rates associated with
close-binary evolution may well tend to produce WR stars with stellar
parameters (e.g., low core mass) most hospitable to dust formation.
On the other hand, efficient dust formation may require both of these
conditions to be simultaneously and independently satisfied: a
short-period binary system for gas-compression and a late-type WR star
with cool temperatures and appropriate chemistry.  If this is the
case, one would expect to find a number of WR(early)+OB binaries with
periods of $\sim$1\,year, a class as-yet-unidentified (e.g., see
Table~3 in Vanbeveren\etal [1998]).

\section{Conclusions}
We report new multi-epoch images of the pinwheel nebula around
WR~98a at 2.2\micron.  By assuming the dust forms 
via colliding winds, orbital parameters of the underlying binary system
have been estimated.
The recent discoveries of pinwheel nebulae around dusty WR stars 
have made possible 
a new approach for identifying and characterizing embedded WR+O
binary systems with $\sim$1~yr periods, systems difficult to identify by any
other means. 
The inferred component separations and high degree of orbital
circularity suggest these systems evolved from red supergiant + OB
binaries, a phase which was terminated by Roche-lobe overflow and the
production of the obscured Wolf-Rayet systems we see today.

\acknowledgements
{
\footnotesize
We acknowledge enlightening discussions with L. Bildsten regarding
circularization time scales and binary stellar evolution.
We would like to thank D. Sivia for the maximum-entropy mapping
program ``VLBMEM,'' and C.D. Matzner, L.J. Greenhill, K. van der Hucht, and
P. Williams
for their helpful comments.
This research has made use of the SIMBAD
database and NASA's ADS Abstract Service.  The data presented herein were
obtained at the Keck Observatory, which is operated as a
scientific partnership among Caltech,
the University of California and NASA.
The Observatory was made possible by the generous
financial support of the W.M. Keck Foundation.  
This work was supported by
the NSF (AST-9315485 \& AST-9731625)
and the ONR (OCNR N00014-97-1-0743-05).}



\begin{thebibliography}{} 

\bibitem[Baldwin et al.\ 1986]{baldwin86}
Baldwin, J. E., Haniff, C. A., Mackay, C. D., \& Warnier, P.J.
1986, Nature, 320, 595

\bibitem[Canto\etal 19996]{crw96}
{Canto}, J., {Raga}, A. C., \& {Wilkin}, F. P. 1996, \apj, 469, 729

\bibitem[Cherchneff \& Tielens 1995]{ct95}
Cherchneff, I. \& {Tielens}, A. G. G. M. 1995, IAU Symposia, 163, 346

\bibitem[Cohen, Kuhi \& Barlow 1975]{cohen75} 
Cohen, M., Kuhi, 
L. V. \& Barlow, M. J. 1975, \aap, 40, 291 

\bibitem[Cohen\etal 1991]{cohen91}
{Cohen}, M., {van der Hucht}, K. A., {Williams}, P. M., \&
        {Th\'e}, P. S. 1991, \apj, 378, 302



\bibitem[Dougherty\etal 1996]{dougherty96}
Dougherty, S. M., {Williams}, P. M., {van der Hucht}, K. A.,
        {Bode}, M. F. \&  {Davis}, R. J. 1996, \mnras, 280, 963


\bibitem[Gull \& Skilling 1984]{gs84}
Gull, S. F. \& Skilling, J. 1984, IEEE Proceedings, 131, Pt. F, No. 6

\bibitem[H\"{o}gbom 1974]{hogbom74}
H\"{o}gbom, J. 1974, \apjs, 15, 417


\bibitem[Le Teuff 1999]{leteuff99}
Le Teuff, Y. H. 1999, personal communication

\bibitem[Marchenko, Moffat \& Grosdidier 1999]{marchenko99} 
Marchenko, S. V., Moffat, A. F. J. \& Grosdidier, Y.  1999, \apj, 522, 433 

\bibitem[Matthews \& Soifer 1994]{ms94}
Matthews, K. \& Soifer, B. T. 1994,   Infrared Astronomy with Arrays:
the Next Generation, I. McLean ed. (Dordrecht: Kluwer Academic Publishers),
p.239

\bibitem[Matthews\etal 1996]{matthews96}
Matthews, K., Ghez, A. M., Weinberger, A. J., \& Neugebauer, G.
1996, \pasp, 108, 615

\bibitem[Moffat\etal 1987]{moffat87}
{Moffat}, A. F. J., {Lamontagne}, R., {Williams}, P. M., Horn, J., \&
{Seggewiss}, W. 1987, \apj, 312, 807

\bibitem[Monnier\etal 1999]{monnier99a} 
Monnier, J. D., Tuthill, 
P. G., Lopez, B., Cruzalebes, P., Danchi, W. C. \& Haniff, C. A. 1999, 
\apj, 512, 351 

\bibitem[Monnier 1999]{monnier99}
Monnier, J. D. 1999, University of California at Berkeley, PhD Dissertation


\bibitem[Sivia 1987]{sivia87}
Sivia, D. S. 1987, Cambridge University, PhD Dissertation

\bibitem[Tassoul 1990]{tassoul90}
Tassoul, J. -L. 1990, \apj, 358, 196

\bibitem[Torres, Conti \& Massey 1986]{tcm86}
{Torres}, A. V., {Conti}, P. S., \& {Massey}, P. 1986, \apj, 300, 379

\bibitem[Tuthill\etal 1999a]{tuthill99a}
Tuthill, P. G., {Monnier}, J. D., \& {Danchi}, W. C.
1999a, \nat, 398, 478

\bibitem[Tuthill et al.\ 1999b]{tuthill99b}
Tuthill, P. G., Monnier, J. D., Danchi, W. C., Wishnow, E., \& Haniff, C. A.
1999b, \pasp, in preparation

\bibitem[Usov 1991]{usov91}
Usov, V. V. 1991, \mnras, 252, 49


\bibitem[van Belle\etal 1999]{vanbelle99}
{van Belle}, G. T., \etal. 1999, \aj, 117, 521

\bibitem[Vanbeveren\etal 1998]{vanbeveren98}
{Vanbeveren}, D., {De Donder}, E., {van Bever}, J.,
{van Rensbergen}, W., \& {De Loore}, C. 1998, New Astronomy, 3, 443

\bibitem[Veen\etal 1998]{veen98}
{Veen}, P. M., {van der Hucht}, K. A., {Williams}, P. M.,
{Catchpole}, R. M., {Duijsens}, M. F. J., {Glass}, I. S.,
\& {Setia Gunawan}, D. Y. A. 1998, \aap, 339, L45


\bibitem[Verbunt \& Phinney 1995]{verbunt95}
{Verbunt}, F. \& {Phinney}, E. S. 1995, \aap, 296, 709

\bibitem[White \& Becker 1995]{wb95}
White, R. L. \& {Becker}, R. H. 1995, \apj, 45, 352

\bibitem[Williams\etal 1987]{williams87}
{Williams}, P. M., {van der Hucht}, K. A., \& {Th\'e}, P. S. 
1987, \aap, 182, 91

\bibitem[Williams\etal 1990]{williams90}
{Williams}, P. M., {van der Hucht}, K. A., {Pollock}, A. M. T.,
{Florkowski}, D. R., {van der Woerd}, H., \& {Wamsteker}, W. M.
1990, \mnras, 243, 662

\bibitem[Williams \& van der Hucht 1992]{wh92}
{Williams}, P. M. \& {van der Hucht}, K. A.,
1992, ASP Conference Series, 22, 269


\bibitem[Williams\etal 1995]{williams95}
{Williams}, P. M.,{Cohen}, M., {van der Hucht}, K. A.,
{Bouchet}, P., \& {Vacca}, W. D. 1995, \mnras, 275, 889

\bibitem[Williams 1998]{williams98}
Williams, P. M. 1998, IAU Symposia, 193, in press

\bibitem[Williams 1999]{williamsPC}
Williams, P. M. 1999, personal communication

\bibitem[Zubko 1998]{zubko98}
Zubko, V. G. 1998, \mnras, 295, 109

\end{thebibliography}
\end{document}